\documentclass[aps,prd,amsfonts, eqsecnum,nofootinbib,twocolumn]
{revtex4} 
\usepackage{graphics,epsfig}
\begin{document}
\preprint{WM-04-116}
%
% Title of paper
\title{Phenomenology of Noncommutative Field Theories\footnote{
Invited plenary talk, IX Mexican Workshop on Particles and Fields,
November 17-22, 2003, Universidad de Colima, Mexico.}
 \vskip 0.1in}
\author{Christopher D. Carone} \email[]{carone@physics.wm.edu}
\affiliation{Particle Theory Group, Department of Physics,
College of William and Mary, Williamsburg, VA 23187-8795}
\date{September 2004}
\begin{abstract}
Experimental limits on the violation of four-dimensional Lorentz
invariance imply that noncommutativity among ordinary spacetime
dimensions must be small.  In this talk, I review the most stringent
bounds on noncommutative field theories and suggest a possible means
of evading them: noncommutativity may be restricted to extra,
compactified spatial dimensions.  Such theories have a number of 
interesting features, including Abelian gauge fields whose Kaluza-Klein
excitations have self couplings.  We consider six-dimensional QED in a
noncommutative bulk, and discuss the collider signatures of the model.
\end{abstract}
\maketitle

%%%%%%%%%%%%%%%%%%%%%%%%%%%%%%%%%%%%%%%%%%%%%%%%%%%%%%%%%%%%%%%%%%%%

\section{Introduction}\label{sec:intro}

%%%%%%%%%%%%%%%%%%%%%%%%%%%%%%%%%%%%%%%%%%%%%%%%%%%%%%%%%%%%%%%%%555
The possibility of extra compactified spatial dimensions at the TeV
scale has led to serious consideration of other modifications of
spacetime structure that may be experimentally accessible.  One such
possibility is that ordinary four-dimensional spacetime may become
noncommutative
~\cite{HPR,HK,Mah,BGXW,GL,ACDNS,MPR,CHK,CST1,ABDG,CCL,CCL4,ccz,RJ} 
at some scale $\Lambda_{{\rm NC}}$:
\begin{equation}\label{eq:com}
\left[\hat{x}^\mu \, , \, \hat{x}^\nu \right]= i \theta^{\mu\nu} \, .
\end{equation}
Here, the position four-vector $x^\mu$ has been promoted to an
operator $\hat{x}^\mu$, and $\theta^{\mu\nu}$ is a real, constant
matrix with elements of order $(\Lambda_{{\rm NC}})^{-2}$.
Noncommutative field theories defined in terms of commuting
coordinates may be constructed by finding an appropriate mapping of
the noncommutative algebra to the space of ordinary
functions~\cite{Madore}.  Given a classical function $f(x)$ with
Fourier transform
\begin{equation}
\tilde{f}(k)=\frac{1}{(2\pi)^{n/2}} \int d^n x \, 
e^{ik_\mu x^\mu} f(x) \,\,\, ,
\end{equation}
one may associate the operator
\begin{equation}
W(f)=\frac{1}{(2\pi)^{n/2}} \int d^n k \,
e^{-ik_\mu \hat{x}^\mu} \tilde{f}(k)
\end{equation}
in the noncommuting theory.  Requiring that this correspondence
holds for the product of functions,
\begin{equation}
W(f) W(g) = W(f \star g)
\end{equation}
one finds that
\begin{equation}\label{eq:mstar}
f \star g = \lim_{y\rightarrow x} e^{\frac{i}{2} \frac{\partial}
{\partial x^\mu}
\theta^{\mu\nu}\frac{\partial}{\partial y^\nu}} f(x) g(y)  \,\,\, .
\end{equation}
This is the Moyal-Weyl $\star$-product~\cite{mwsp}.  To understand its
usefulness, notice that
\begin{eqnarray}
\left[x^\mu\stackrel{\star}{,} x^\nu\right] &=& x^\mu (1+\frac{i}{2}
\stackrel{\leftarrow}{\partial}\cdot\theta\cdot\stackrel{\rightarrow}
{\partial})
x^\nu - (\nu \leftrightarrow \mu) \nonumber \\ & = & \frac{i}{2}
\theta^{\mu\nu}-\frac{i}{2} \theta^{\nu\mu} = i \theta^{\mu\nu} \,\,\, .
\end{eqnarray}
The star product allows one to reproduce the original operator algebra
while working exclusively with ordinary functions.

The star product represents the starting point for the phenomenological
study of noncommutative field theories.  For example, one can immediately
write down the noncommutative generalization of $\lambda \phi^4$ theory,
\begin{equation}
{\cal L} =  \frac{1}{2} \partial_\mu \phi
\partial^\mu \phi-\frac{1}{2}m^2 \phi^2 - \frac{\lambda}{4\!}  
(\phi\star\phi)^2 \, ,
\end{equation}
where we have used 
\begin{equation}
\int d^4 x \, f\star g = \int d^4 x \, f \, g \,\,\, ,
\end{equation}
which follows from integration by parts.   
Noncommutative gauge theories require star multiplication as
well as a modified form of the Lagrangian in order to preserve
the desired local symmetries of the theory.  For example, the
Lagrangian of noncommutative QED (NCQED) is given by~\cite{HAY}
\begin{equation}
{\cal L} = -\frac{1}{4} F_{\mu\nu} \star F^{\mu\nu}
+ \overline{\psi}\star (i\not\!\!D-m)\star \psi  \,\,\, ,
\label{eq:ncqedord}
\end{equation}
\begin{equation}
D_\mu=\partial_\mu-i e A_\mu \,\,\, ,
\label{eq:covder}
\end{equation}
\begin{equation}
F_{\mu\nu}=\partial_\mu A_\nu - \partial_\nu A_\mu -
i e \, [A_\mu\stackrel{\star}{\,,}A_\nu] \,\,\, ,
\label{eq:fmunu}
\end{equation}
which is invariant under the noncommutative U(1) gauge transformation
\begin{equation}
\psi(x) \rightarrow \psi'(x) = U \star \psi(x),
\end{equation}
\begin{equation}
A_\mu(x) \rightarrow A'_\mu(x)= U\star
A_\mu(x) \star U^{-1} + \frac{i}{e}\, U \star \partial_\mu U^{-1} .
\end{equation}
The gauge transformation matrix $U$ is a star exponential, $(e^{i
\alpha})_\star$, in which each occurrence of ordinary multiplication in
the the Taylor expansion of $e^{i \alpha}$ is replaced by a star
product.  The Lagrangian in Eqs.(\ref{eq:ncqedord})-(\ref{eq:fmunu}) is
similar in form to that of an ordinary non-Abelian gauge theory.  In
particular, the gauge boson self interactions that originate from the
noncommutativity of the group generators in a non-Abelian theory arise
in NCQED due to the noncommutativity of ordinary functions under star
multiplication.

Phenomenological studies of noncommutative gauge theories have
focused primarily on the collider~\cite{HPR,HK,Mah,BGXW,GL}
and low-energy signatures~\cite{MPR,CHK,CST1,ABDG,CCL,CCL4} of NCQED.
Provided that the scale $\Lambda_{{\rm NC}}$ is sufficiently low,
the new photon self-interactions in Eq.~(\ref{eq:ncqedord}) may
be discerned at, for example, the Next Linear Collider 
(NLC)~\cite{HPR}. On the other hand,
the most striking phenomenological feature of noncommutative theories is that
they are Lorentz violating. The constant parameter $\theta^{\mu\nu}$ 
defines the preferred directions $\theta^{0i}$ and 
$\epsilon^{ijk}\theta^{jk}$ in a given Lorentz frame.  Low-energy tests 
of Lorentz invariance~\cite{lv} place bounds on $\Lambda_{{\rm NC}}$ of order 
$10$~TeV, if one considers NCQED processes at tree-level~\cite{CHK}.  
However, there are operators generated at one- and two-loops
that are more stringently bounded~\cite{ABDG,CCL,CCL4}.  As pointed 
out by Anisimov, Banks, Dine and Graesser~\cite{ABDG}, effective 
interactions such as
\begin{equation}
\begin{array}{cc}
{\cal O}_1 = m_e \theta^{\mu\nu} \bar\psi \sigma_{\mu\nu} \psi &
{\cal O}_2 = \theta^{\mu\nu} \bar\psi D_\mu \gamma_\nu \psi  \\
{\cal O}_3 = (\theta^2)^{\mu\nu} F_{\mu\rho} F^\rho_\nu &
{\cal O}_4 = \theta^{\mu\nu} \theta^{\rho\sigma} F_{\mu\nu} F_{\rho\sigma}
\end{array}
\end{equation}
are constrained by a variety of low-energy and astrophysical processes.
Notable, the operator ${\cal O}_1$ will affects a spin-polarized torsion 
pendulum by providing a coupling between the net spin, and the fixed 
external ``B field'' defined by $\theta^{ij}$.  In NCQED, this operator is 
generated via the diagrams shown in Fig.~1

%%%%%%%%%%%%%%%%%%%%%%%%%%%%%%%%%%%%%%%%%%%
\vspace{1em}
\centerline{\epsfxsize 3.1in \epsfbox{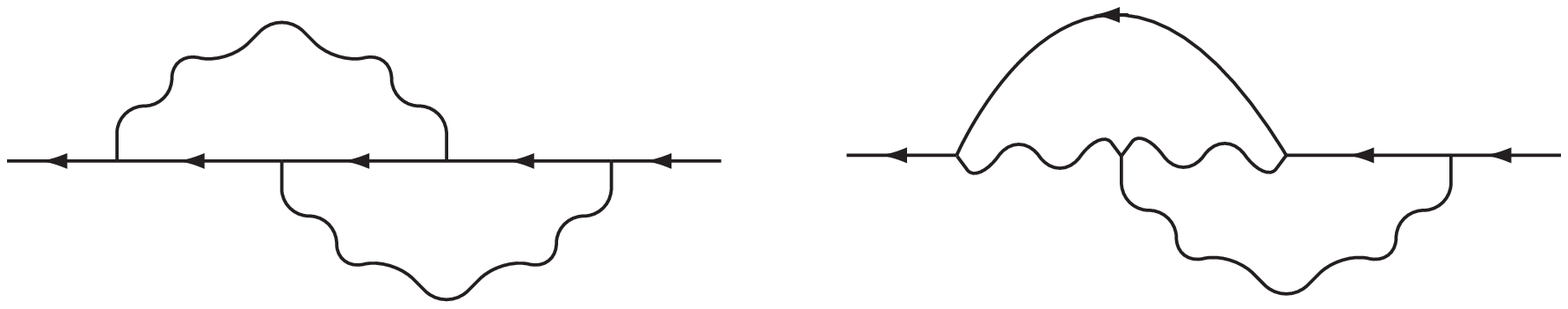}}
\centerline{Fig.~1 Two-loop diagrams that contribute to ${\cal O}_1$.}
\vspace{1em}
%%%%%%%%%%%%%%%%%%%%%%%%%%%%%%%%%%%%%%%%%%

Integrating over the full range of loop momenta, the diagrams in Fig.~1
yield the following finite amplitude $A$:
\begin{eqnarray}
    A & = & 24 i m_e e^4 \int \frac{d^4 k}{(2\pi)^4} \frac{d^4 l}{(2\pi)^4} \,
\frac{ e^{i\, l\cdot\theta\cdot k} \sigma_{\mu\nu} k^\mu l^\nu}
{{k^2 l^4 (k+l)^4}}   \nonumber \\
&& =  \frac{1}{8} m_e \alpha^2
    \frac{\sigma_{\mu\nu} \theta^{\mu\nu}}{
\sqrt{(-1/2){\rm Tr}\, \theta^2} }
\label{eq:weird}
\end{eqnarray}
Notably, the peculiar form of this result implies that the coefficient
of operator ${\cal O}_1$ is $\sim \frac{1}{8} m_e \alpha^2 \sim
10^{-5}$~MeV even when $\Lambda_{NC}$ is taken to be arbitrarily
large. If this result is taken at face value, then one concludes that
noncommutative field
theories are ruled out:  the experimental upper bound on the same operator
coefficient is $O(10^{-25})$~MeV~\cite{ABDG}.  However, the appearance of
a $\theta^{\mu\nu}$ in the denominator of Eq.~(\ref{eq:weird}) is a
consequence of the high-momentum region of integration, and may not occur
if the physics is altered at a high scale.  In particular, the same
integral may be evaluated with an ultraviolet cutoff $\Lambda$, yielding
\begin{equation}
A_\Lambda = \frac{3}{4} m_e \Lambda^2 \left(\frac{\alpha^2}{4\pi}\right)^2
\sigma_{\mu\nu} \theta^{\mu\nu} \, ,
\end{equation}
which leads to the bound $\theta\Lambda^2 < 10^{-19}$.  The
implications of this result are no less striking: $\theta^{\mu\nu}$
must be $19$ orders of magnitude smaller than the size one might
expect based on naive dimensional arguments in order that the theory
remain consistent with the experimental data.  One may worry that this
simple estimate involves a cut off that violates the underlying gauge 
invariance of the theory.  However, studies of softly-broken supersymmetric
noncommutative QED lead to qualitatively similar results, with the
scale of supersymmetry breaking providing a natural gauge-invariant 
regulator~\cite{CCL4}.

%%%%%%%%%%%%%%%%%%%%%%%%%%%%%%%%%%%%%%%%%
\renewcommand{\theequation}{2.\arabic{equation}}
\setcounter{equation}{0}
\section{Noncommutative Bulk}

%%%%%%%%%%%%%%%%%%%%%%%%%%%%%%%%%%%%%%%%%

A possible way of avoiding the stringent bounds on the violation of
four-dimensional (4D) Lorentz invariance is to restrict the
noncommutativity to extra spatial dimensions~\cite{CCXD,GMW}. (For another 
approach, see Ref.~\cite{ccz}.)  The simplest possibility
is six-dimensional (6D) QED with noncommutativity restricted to the
fifth and sixth dimensions~\cite{CCXD}.  While both the
compactification of the extra dimensions and the noncommutativity
break 6D Lorentz invariance, 4D Lorentz invariance remains intact.

We consider a model with gauge fields defined on the full space and
fermion fields restricted to a brane~\cite{CCXD}.  The Lagrangian is
\begin{eqnarray}
{\cal L}_6 &=& - \frac{1}{4} {\cal F}_{MN} \star {\cal F}^{MN}
         + {\cal L}_{gauge\ fixing}
                              \nonumber \\
        &+& \delta^{(2)}(\vec y) \left\{
           \bar\psi (i\not\!\partial -m) \psi
         + \hat e \bar\psi \star \not\!\!\! {\cal A} \star \psi
                                          \right\}  \ ,
\end{eqnarray}

\noindent where
\begin{equation}
{\cal F}_{MN} = \partial_M {\cal A}_N - \partial_N {\cal A}_M
               -i \hat e \left[{\cal A}_M
                 \stackrel{\star}{,} {\cal A}_N
                                                \right]  \ ,
\end{equation}

\noindent and where $\hat{e}$ is the 6D gauge coupling.  Our notation for 
the position six-vector is
\begin{equation}
X^M  = (x^0,x^1,x^2,x^3,y^5,y^6)  \ ,
\end{equation}

\noindent with $\vec y \equiv (y^5,y^6)$.

We compactify the extra dimensions on the orbifold $T^2/ \mathbb Z_2$,
where $T^2$ is a general 2-torus.  We take into account the
possibility of two different radii $R_5$ and $R_6$ and a relative
shift angle $\phi$ between the two directions of
compactification~\cite{kd}.  This allows us to avoid zeros in the
scattering amplitudes of interest to us, that appear only in the limit
$R_5=R_6$ and $\phi=\pi/2$.  The coordinates along the torus are
$\zeta^i$, related to $y^i$ by
\begin{eqnarray}
y^5 &=& \zeta^5 + \zeta^6 \cos\phi  \nonumber \\
y^6 &=& \zeta^6 \sin\phi           \ .
\end{eqnarray}

\noindent The periodicity requirements on a function of orbifold
coordinates $f(\zeta^5,\zeta^6)$ are
\begin{equation}\label{eq:reqs}
f(\zeta^5,\zeta^6) = f(\zeta^5 + 2\pi R_5, \zeta^6)
= f(\zeta^5, \zeta^6 + 2\pi R_6)    \ .
\end{equation}

Without orbifolding, Eq.~(\ref{eq:reqs}) implies that bulk fields have
6D wave functions proportional to
\begin{eqnarray}
&&\exp \left\{ i \frac{n^5 \zeta^5}{ R_5 } +i \frac{n^6 \zeta^6}{ R_6 }
\right\}
               \nonumber \\
&=& \exp  \left\{ i \frac{n^5 y^5}{ R_5} + i \frac{y^6}{\sin\phi}
             \left[ \frac{n^6}{ R_6} - \frac{n^5}{R_5}\cos\phi \right]
                                                \right\} \ ,
\end{eqnarray}

\noindent  where $n^5$ and $n^6$ are integers. The masses of the KK modes
are eigenvalues of the mass operator
$-\partial^2_{y^5}-\partial^2_{y^6}$ and are given by
\begin{equation}
m_{\vec n}^2 = \frac{1}{ \sin^2 \phi} \left( \frac{n_5^2}{ R_5^2}
                                        + \frac{n_6^2}{R_6^2}
                  - \frac{2 n_5 n_6}{R_5 R_6} \cos\phi   \right)  \ ,
\end{equation}

\noindent where $\vec n \equiv (n^5,n^6)$.

The $\mathbb Z_2$ orbifolding consists of identifying points connected
by $\vec{y} \rightarrow -\vec{y}$.  Different
components of the gauge field may be Fourier expanded with
different $\mathbb Z_2$-parities so that zero modes are only present
for the first four components,
\begin{equation}
{\cal A}_M(X) = \sum_{\{\vec n_+\}} {\cal A}_M^{(\vec n)} (x) \, 
f_{\vec{n}}(\zeta^5,\zeta^6) \,\,\, ,
\end{equation}
where
\begin{eqnarray}
f_{\vec{n}}(\zeta^5,\zeta^6) =
\left\{
  \begin{array}{ll}
    \cos \left( \frac{n^5 \zeta^5 + \xi n^6 \zeta^6 }{ R }\right)\, ,  &
 M = \mu
                            \\[2ex]
    \sin \left( \frac{n^5 \zeta^5 + \xi n^6 \zeta^6}{ R }\right)\, ,  &
 M = 5,6
  \end{array}   \right.
\end{eqnarray}

\noindent with $\mu=0,1,2,3$. Here, $\xi$ is the ratio of the radii,
with
\begin{equation}
R_5 \equiv R = \xi R_6   \ .
\end{equation}
Since orbifolding has provided wave functions with distinct parities, the
value of $\vec{n}$ is now restricted to a half plane
including the origin,
\begin{eqnarray}
{\{\vec n_+\}} = \left\{
  \begin{array}{l}
     \vec n = \vec 0; \quad {\rm or} \\
     n^5 = 0, n^6 > 0; \quad {\rm or} \\
     n^5 > 0, n^6 = {\rm any\ integer}
  \end{array}                              \right\} \ ,
\end{eqnarray}

\noindent  Within the set $\{\vec n_+\}$, masses are unique for 
most values of
$\xi$ and $\phi$.

One obtains the 4D Lagrangian by integrating over the extra dimensions,
\begin{equation}
{\cal L}_4 = \int d^2 y \, {\cal L}_6
           = \frac{1}{\xi_0} \int_0^{2\pi R} d\zeta^5 \,
             \int_0^{2\pi R} d\tilde{\zeta^6} \,   {\cal L}_6 \ ,
\end{equation}

\noindent where $\xi_0 \equiv \xi / \sin\phi$ and
$\tilde{\zeta^6}\equiv \xi \zeta^6$.

The gauge fixing Lagrangian is chosen as
\begin{equation}
{\cal L}_{gauge\ fixing} = -\frac{1}{2} \eta
        \left( \partial_\mu {\cal A}^\mu
              + \frac{1}{\eta} \partial_k {\cal A}^k  \right)^2
\end{equation}

\noindent with $k$ = 5,6. Terms in the Lagrangian quadratic in the
gauge field become,
\begin{eqnarray}
&&\left(  -\frac{1}{4} {\cal F}_{MN} {\cal F}^{MN}
       + {\cal L}_{gauge\ fixing}  \right)_{free,\ 4d} \nonumber \\[2ex]
   &=& -\frac{1}{4} F_{\mu\nu}^{(\vec 0)} F^{\mu\nu (\vec 0)}
     -\frac{1}{2} \eta \left( \partial^\mu A_\mu^{(\vec 0)} \right)^2
                                               \nonumber  \\
&+&{\sum}' \Bigg\{ -\frac{1}{ 4}  F_{\mu\nu}^{(\vec n)} F^{\mu\nu (\vec n)}
        -\frac{1}{2} \eta \left( \partial^\mu A_\mu^{(\vec n)} \right)^2
                                                \nonumber          \\
&+& 
      \frac{1}{2} \partial_\mu A_L^{(\vec n)}\partial^\mu A_L^{(\vec n)}
      +\frac{1}{2} \partial_\mu A_H^{(\vec n)}\partial^\mu A_H^{(\vec n)}
                                              \nonumber \\
&+& \frac{1}{2} m_{\vec n}^2    A^{\mu (\vec n)}   A_\mu^{(\vec n)}  
- \frac{1}{2} m_{\vec n}^2    A_L^{(\vec n)}   A_L^{(\vec n)}
                                               \nonumber \\
&-& \frac{1}{ 2\eta} m_{\vec n}^2  A_H^{(\vec n)}   A_H^{(\vec n)}
                             \Bigg\}.   
\end{eqnarray}

\noindent The primed sum is over the Kaluza-Klein (KK) modes, i.e., over
$\{\vec n_+\}$ excluding $\vec 0$.  The 6D fields $\cal A$ have been
rescaled,
\begin{eqnarray}
{\cal A}_M^{(\vec 0)} &=&  \frac{ \sqrt{\xi_0}}{2\pi R} A_M^{(\vec 0)} \ ,
                                        \nonumber \\
{\cal A}_M^{(\vec n)} &=&  \frac{ \sqrt{2\xi_0}}{ 2\pi R} A_M^{(\vec n)}
                         \qquad  [\vec n \not= \vec 0] \ ,
\end{eqnarray}

\noindent where the fields $A$ have their canonical 4D mass dimensions.
The fifth and sixth components have been combined into
\begin{eqnarray}
A_L^{(\vec n)} = \frac{1}{ |\vec {\tilde n}|}
           \left( \tilde n^5 A^{6 (\vec n)} - \tilde n^6 A^{5 (\vec n)}
                                  \right)  \ ,
                                            \nonumber \\
A_H^{(\vec n)} = \frac{1}{ |\vec {\tilde n}|}
           \left( \tilde n^5 A^{5 (\vec n)} + \tilde n^6 A^{6 (\vec n)}
                                            \right)  \ ,
\end{eqnarray}

\noindent where $\vec {\tilde n} =(\tilde n^5, \tilde n^6)$ with
\begin{eqnarray} \label{eq:ntilde}
\tilde n^5 &=& n^5  \nonumber \\
\tilde n^6 &=& \frac{1}{ \sin\phi}
            \left( \xi n^6 - n^5 \cos\phi \right)  \ ,
\end{eqnarray}

\noindent  and $m_{\vec n} = |\vec{\tilde n}|/R$.  The fields
$A_L$ and $A_H$ are physical and unphysical scalars in the 4D
theory, respectively. As $\eta \rightarrow 0$, the field $A_H$ is removed
from the theory, the extra-dimensional generalization of unitary gauge.
We work in the $\eta \rightarrow 0$ limit henceforth.  Thus, from the
free gauge Lagrangian the physical states are the
ordinary massless photon, the vector KK modes, and the scalar KK modes
$A_L^{(\vec n)}$.

The fermion fields $\psi$ are defined only at the $\vec{y}=\vec{0}$ orbifold
fixed point, and involve no rescaling.  Since $A_L^{(\vec{n})}$ is odd under
the $\mathbb Z_2$ parity it vanishes at $\vec{y}=\vec{0}$.
Hence the fermions interact only with the photon and its vector KK
excitations.  The fermion Lagrangian is
\begin{eqnarray}
{\cal L}_{f,4d} &=& \bar\psi (i\not\! \partial - m) \psi
                + e \bar\psi \, \star \not\!\! A^{(0)} \star \psi
                                          \nonumber \\
                &+& e \sqrt{2} \, {\sum}' \,
                  \bar\psi \, \star \not\!\! A^{(\vec n)} \star \psi  \ .
\label{eq:branelag}
\end{eqnarray}

\noindent The 4D gauge coupling has been identified through the rescaling
\begin{equation}
\hat e = \frac{2\pi R}{ \sqrt{\xi_0}} e \ .
\end{equation}

The pure gauge field interactions come from the terms
\[
{\cal L}_{g\ int,6d} =  i \hat e  \partial_M {\cal A}_N
         \left[{\cal A}^M  \stackrel{\star}{,} {\cal A}^N  \right]
\hspace{10em}\]\begin{equation}
+ \frac{1}{ 4} \hat e^2
         \big[{\cal A}_M  \stackrel{\star}{,} {\cal A}_N  \big] \star
         \big[{\cal A}^M  \stackrel{\star}{,} {\cal A}^N  \big]  \ ,
\end{equation}

\noindent in which the Moyal commutator may be written as
\[
\left[{\cal A}_M  \stackrel{\star}{,} {\cal A}_N  \right] = \hspace{18em}
\]
\begin{equation}
2i \lim_{X \rightarrow Y}
           \sin\left(\frac{1}{ 2}
  \frac{\partial}{\partial X^i} \theta^{ij} \frac{\partial}{\partial Y^j}
                  \right){\cal A}_M(X) {\cal A}_N(Y)  \ .
\end{equation}
One may now extract the three-photon coupling in the 4D Lagrangian,
\[
{\cal L}_{3\gamma,4d} = -e\sqrt{2} \, {\sum}'
       \left( \delta_{\vec n_a, \vec n_b + \vec n_c}
          +   \delta_{\vec n_b, \vec n_c + \vec n_a}
          -   \delta_{\vec n_c, \vec n_a + \vec n_b}  \right)
\]\begin{equation}                                      
  \times \partial_\alpha A_\beta^{(\vec n_c)}
           A^{\alpha (\vec n_a)} A^{\beta (\vec n_b)}
  \sin\left(\frac{\tilde n_a^i \theta_{ij} \tilde n_b^j}{2R^2} \right) \ ,
\end{equation}

\noindent where $\tilde{n}$ is defined in Eq.~(\ref{eq:ntilde}).
The triple-photon couplings involve only the KK modes, and never any
ordinary massless photons.  The Feynman rule that corresponds to the
$3\gamma$ term in the Lagrangian, for the momenta, Lorentz indices, and
KK modes labeled in Fig.~2, is given by
\begin{eqnarray}\label{eq:thevertex}
V_{3\gamma} &=& -e\sqrt{2}
       \left( \delta_{\vec n_a, \vec n_b + \vec n_c}
          +   \delta_{\vec n_b, \vec n_c + \vec n_a}
          -   \delta_{\vec n_c, \vec n_a + \vec n_b}  \right)
                                      \nonumber \\
                        &\times&
   \sin\left(\frac{\tilde n_a^i \theta_{ij} \tilde n_b^j}{2R^2} \right)
                                                \\
                        &\times&
  \left[ g_{\mu\nu} (p-q)_\rho +g_{\nu\rho} (q-r)_\mu
         +  g_{\rho\mu} (r-p)_\nu    \right]    \ .  \nonumber
\end{eqnarray}

%%%%%%%%%%%%%%%%%%%%%%%%%%%%%%%%%%%%%%
\vspace{1em}
\centerline {\epsfxsize 2.0in \epsfbox{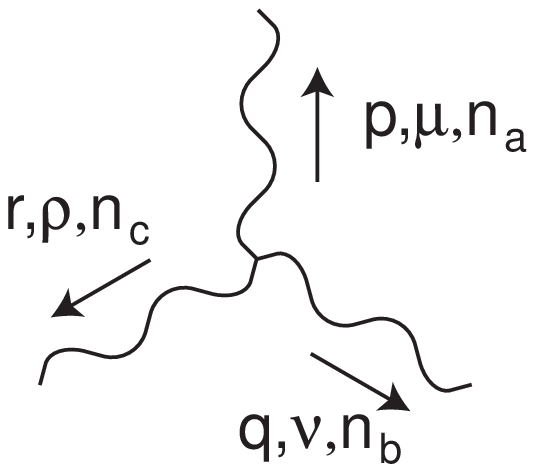}}
\centerline {Fig. 2. The triple KK photon vertex.}
\vspace{1em}
%%%%%%%%%%%%%%%%%%%%%%%%%%%%%%%%%%%%%%

When the noncommutativity is only in the
extra dimensions, the only independent non-zero component of the
noncommutativity tensor is $\theta^{56} \equiv \theta$.
(Theories with space-like noncommutativity are known to preserve
perturbative unitarity~\cite{gomis}.) The argument of
the sine simplifies using
\begin{equation}
\tilde n_a^i \theta_{ij} \tilde n_b^j = \xi_0 \theta
              \left(n_a^5 n_b^6 - n_a^6 n_b^5 \right) \ .
\end{equation}

A four-photon vertex may be computed in a similar way, but will
not be relevant to the physical processes studied in the sections
that follow.
%%%%%%%%%%%%%%%%%%%%%%%%%%%%%%%%%%%%%%%%%%%%%%%%%%%%%%%%
\renewcommand{\theequation}{3.\arabic{equation}}
\setcounter{equation}{0}
\section{U(1) Charges}
%%%%%%%%%%%%%%%%%%%%%%%%%%%%%%%%%%%%%%%%%%%%%%%%%%%%%%%%

Let us now focus on the couplings of the gauge field to matter
localized on the $\vec{y}=0$ brane. Since the field 
$\psi \equiv \psi(x^\mu,\vec{0})$ is independent of
the coordinates $y^5$ and $y^6$, the star products in
Eq.~(\ref{eq:branelag}) reduce trivially to ordinary multiplication.
At this point, one might suspect that there is no restriction on
the allowed U(1) charges for brane-localized matter.  To check
the consistency of this claim, consider the behavior of gauge 
transformations near the $\vec{y}=0$ brane.  Recall, that the full 
6D noncommutative gauge transformation is given by
\begin{equation}
A \rightarrow U \star A \star U^{-1}
+ \frac{i}{e}\, U \star \partial_\mu U^{-1} .
\label{eq:6dncgt}
\end{equation}
Infinitesimally, this may be written
\begin{equation}
\delta A_\mu = \frac{1}{e} \partial_\mu \alpha + i [\alpha
\stackrel{\star}{,} A_\mu ] + \cdots \,\,\, .
\end{equation}
Since $A^\mu$, and hence $\delta A^\mu$, are even under the $Z_2$ orbifold
parity, it follows that the gauge parameter $\alpha$ is also an even function.
Thus,
\begin{equation}
\frac{\partial}{\partial y^5} \alpha |_{\vec{y}=0}
=\frac{\partial}{\partial y^6} \alpha |_{\vec{y}=0} =0 \,\,\, ,
\end{equation}
and Eq.~(\ref{eq:6dncgt}) reduces to the familiar result
\begin{equation}
A^\mu (x^\mu, \vec{0}) \rightarrow A^\mu (x^\mu, \vec{0})
+ \frac{i}{e} \partial^\mu
\end{equation}
on the $\vec{y}=0$ brane.  Gauge invariance places no restriction on
the U(1) charges of matter on the brane~\cite{CCXD}, unlike the case 
in 4D NCQED~\cite{HAY}.

%%%%%%%%%%%%%%%%%%%%%%%%%%%%%%%%%%%%%%
\vspace{1em}
\centerline{\epsfxsize 3in \epsfbox{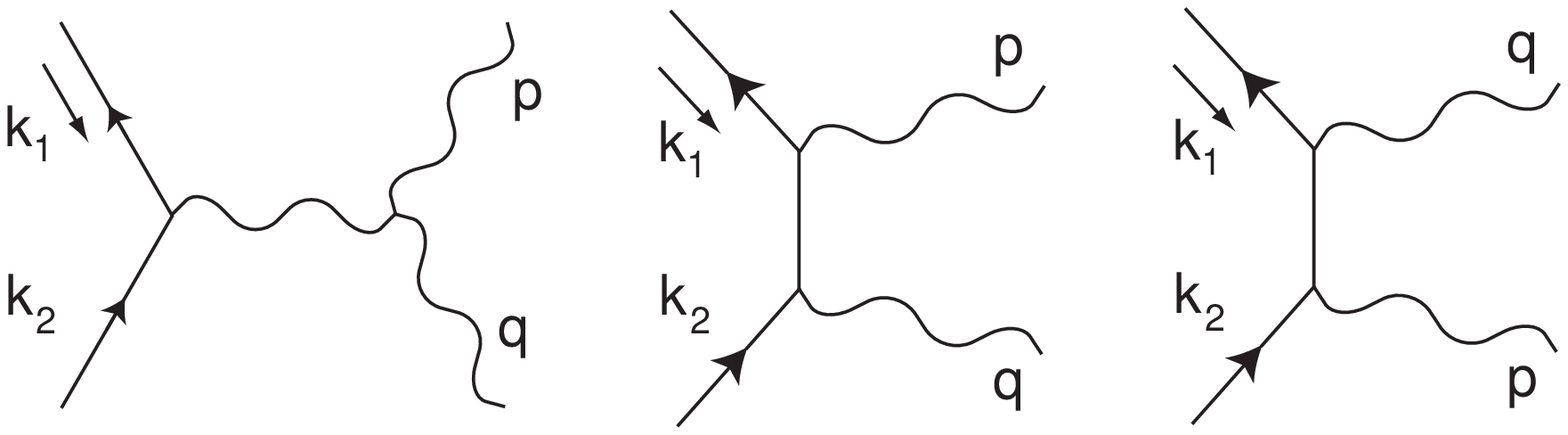}}
Fig. 3. Feynman diagrams for production of KK
pairs, including the noncommutative triple photon vertex.

%%%%%%%%%%%%%%%%%%%%%%%%%%%%%%%%%%%%%%

%%%%%%%%%%%%%%%%%%%%%%%%%%%%%%%%%%%%%%%%%%%%%%%%%%%%%%%%
\renewcommand{\theequation}{4.\arabic{equation}}
\setcounter{equation}{0}
\section{The Smoking Gun}
%%%%%%%%%%%%%%%%%%%%%%%%%%%%%%%%%%%%%%%%%%%%%%%%%%%%%%%%
Pair production of KK photons at colliders can occur through the
Feynman diagrams shown in Fig.~3.  Notice that the first diagram
involves the noncommutative triple-photon vertex, Eq.~(\ref{eq:thevertex}).

We present the cross section for the case that the Kaluza-Klein
states are the $\vec n = (1,0)$ and $(1,1)$ modes.  While noncommutativity
also affects the production of $(0,1)$-$(1,0)$ pairs, the nonstandard
diagram in this case involves contributions from different intermediate
states that tend to cancel, suppressing the rate.  At the parton level,
for the final states we have chosen, the nonstandard cross section is
\[
\sigma_{NS} (\hat s) = \frac{\pi\alpha^2\lambda}{3 \hat s^2 } 
    \left[ \frac{(m_{21}^2 - m_{01}^2)/(m_{11}m_{10})}{
                  (\hat{s}-m_{21}^2)(\hat{s}-m_{01}^2)} \right]^2
           \!\! \sin^2 \left( \frac{\xi_0\theta}{2R^2} \right) \]
\[                              \times
\Big\{ \hat s^4 +8(m_{11}^2+m_{10}^2) \hat s^3
      -[18m_{11}^4+32m_{11}^2 m_{10}^2+18 m_{10}^4] \hat s^2
                                      \]
\[                              
+ 8 (m_{11}^6-4m_{11}^4 m_{10}^2-4m_{11}^2 m_{10}^4+m_{10}^6) \hat s
                                      \]
\begin{equation}                            
+ ( m_{11}^2-m_{10}^2)^2 (m_{11}^4 +10m_{11}^2 m_{10}^2 +m_{10}^4)
\Big\}     \ ,
\end{equation}

\noindent where $\hat s$ is the partonic center-of-mass (CM) energy
squared and $\lambda$ is defined in terms of the CM 3-momentum of
either final state particle, $|\vec p| = \lambda/2\sqrt{\hat s}$ with
\begin{equation}
\lambda = \sqrt{\hat s^2 - 2 \hat s (m_{11}^2+m_{10}^2) +
(m_{11}^2-m_{10}^2)^2} \ .
\end{equation}

\noindent The
initial partons are treated as massless.  The parton level cross section
for the standard  process is
\[
\sigma_{S}(\hat s)  = \frac{16\pi\alpha^2}{ \hat s^2}
   \Bigg\{ -\lambda \frac{m_{11}+m_{10}}{ m_{11}}
\]\begin{equation}                       
+ \frac{\hat s^2 +\left( m_{11}^2+m_{10}^2 \right)^2 }{ \hat s - m_{11}^2
-m_{10}^2 }
  \ln\frac{ \hat s-m_{11}^2-m_{10}^2 + \lambda }{ \hat s-m_{11}^2-m_{10}^2
- \lambda }                                           \Bigg\}  \ .
\end{equation}

The collider cross section is
\[
\sigma (s,AB \rightarrow \gamma_{11} \gamma_{10} X)
  = \int_\tau^1 dx_1 \int_{\tau/x_1}^1 dx_2
\]\[
\times
\frac{1}{3} \sum_q \left[ f_{q/A}(x_1) f_{\bar q/B}(x_2)
                + f_{\bar q/A}(x_1) f_{q/B}(x_2) \right]
\]
\begin{equation}\times
\left\{ e_q^4 \sigma_{S}(\hat s) + e_q^2 \sigma_{NS}(\hat s) \right\}
\ ,
\end{equation}

\noindent where $\hat s = x_1 x_2 s$, $\tau = \hat s_{min}/s$, and
$\hat s_{min}$ is the square of the sum of the KK excitation
masses, or
\begin{equation}
\hat s_{min} = \left(  m_{10}+m_{11}  \right)^2  \ .
\end{equation}

\noindent Also, $f_{q/A}(x) = f_{q/A}(x,\mu)$ are the parton distribution
functions for quark $q$ in hadron $A$ evaluated at renormalization scale
$\mu$, and the $1/3$ is from color averaging.

We evaluated the cross section for a proton-proton collider using the
CTEQ5L parton distribution functions~\cite{5L} at a fixed scale $\mu =
2$ TeV, with $\xi = 0.8$, $\phi=1.5$ and $\sin^2(\xi_0 \theta/2
R^2)=1$ .  Fig.~4 shows the event rate over a range of center of mass
energies for $1/R=4$~TeV and $100$~fb$^{-1}$ of integrated luminosity.
At a $200$~TeV VLHC, for example, we find $294$ events where there is
an expectation without noncommutativity of $165 \pm 12.8$, a $10.1$
sigma effect.  This is a significant signal for a choice of $1/R$ that
is consistent with current indirect bounds on the compactification
scale~\cite{pew}. What is more significant is the relative effect of
the noncommutative vertex Eq.~(\ref{eq:thevertex}) on the production
of different KK mode pairs.  For example, production of
$(1,0)$-$(1,0)$ pairs receives {\em no} noncommutative corrections
while production of $(1,0)$-$(1,1)$ pairs does.  Comparison of these
channels may help eliminate systematic uncertainty originating, for
example, from parton distribution functions.

%%%%%%%%%%%%%%%%%%%%%%%%%%%%%%%%%%%%%%
\vspace{1em}
\centerline{\epsfxsize 3in \epsfbox{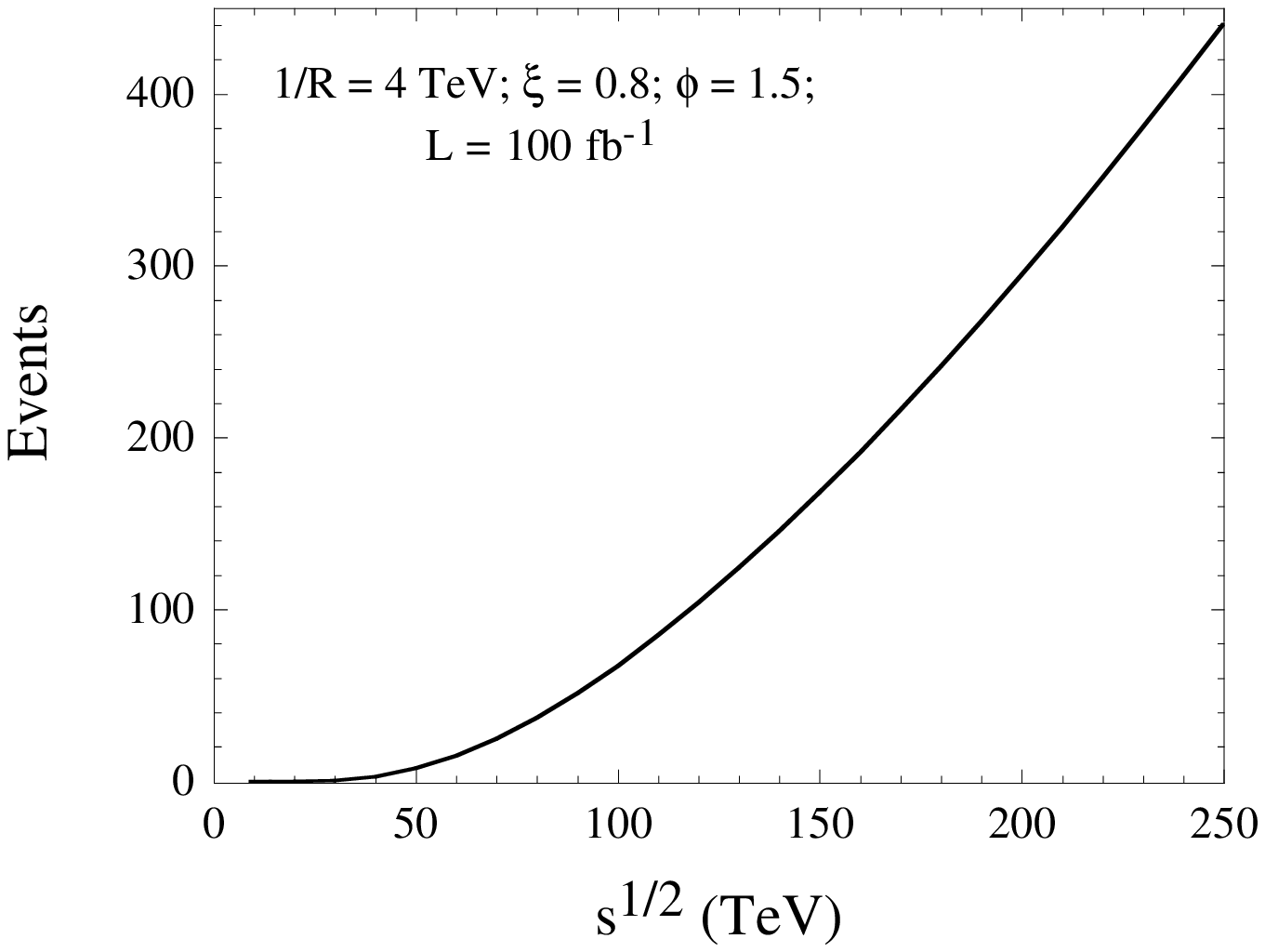}}
Fig. 4. Event rate at a proton-proton collider for the production of
the KK pair $\gamma_{11}+\gamma_{10}$ vs. the $pp$ center-of-mass energy
$\sqrt{s}$.
\vspace{1em}
%%%%%%%%%%%%%%%%%%%%%%%%%%%%%%%%%%%%%%

%%%%%%%%%%%%%%%%%%%%%%%%%%%%%%%%%%%%%%%%%%%%%%%%%%%%%%%
\section{Conclusions}
\renewcommand{\theequation}{5.\arabic{equation}}
\setcounter{equation}{0}
%%%%%%%%%%%%%%%%%%%%%%%%%%%%%%%%%%%%%%%%%%%%%%%%%%%%%%

Four-dimensional noncommutative theories are tightly constrained by
low-energy searches for the violation of Lorentz-invariance.  By
restricting noncommutativity to the bulk, we avoid conflict with the
most stringent experimental limits, which otherwise force the
magnitude of noncommutativity to be small.  We presented an explicit
example, based on the orbifold $T^2/ \mathbb Z_2$, to illustrate the
effects of spatial noncommutativity in 6D QED with fermions confined
to an orbifold fixed point.  Notably, we find new three- and
four-point couplings involving KK excitations of the photon, but not
its zero mode.  With extra dimensions at the TeV scale, the most
promising way of detecting these interactions is through the pair
production of KK modes at a very large hadron collider (VLHC), $f
{\bar f}\rightarrow \gamma^{(\vec{m})}\gamma^{(\vec{n})}$, with
$\vec{m} \neq \vec{n}$. Observing order 100\% corrections to the
production of certain pairs of KK modes at a VLHC while finding no
corrections to others would provide a clear signal of noncommutativity
in the bulk.
%%%%%%%%%%%%%%%%%%%%%%%%%%%%%%%%%%%%%%%%%%%%%%%%%%%%%%%%%%%%%%%%%%%%%%%%

%\appendix
%\section{}

%%%%%%%%%%%%%%%%%%%%%%%%%%%%%%%%%%%%%%%%%%%%%%%%%%%%%%%%%%%%%%%%%%%%%%%%

%%%%%%%%%%%%%%%%%%%%%%%%%%%%%%%%%%
\begin{acknowledgments}
CDC thanks the NSF for support under Grant 
No.~PHY-0140012 and supplement PHY-0352413. In addition, CDC thanks
Alfredo Aranda, Paulo Amore and the Physics Department at the 
University of Colima for their hospitality.
\end{acknowledgments}

%\bibliography{}

\end{document}